\documentclass[aps,prl,twocolumn,showpacs,showkeys,superscriptaddress]{revtex4}
\usepackage{ucs}
\usepackage[warn]{mathtext}
\usepackage[utf8x]{inputenc}
\usepackage[xdvi]{graphicx}
\usepackage{amssymb,amsmath}
\begin{document}

\title{Pressure effect on magnetic susceptibility of LaCoO$_3$}

\author{A.S. Panfilov}
\email[]{panfilov@ilt.kharkov.ua}
\affiliation{B. Verkin Institute for Low Temperature Physics and Engineering,
National Academy of Sciences of Ukraine, 61103 Kharkov, Ukraine}

\author{G.E. Grechnev}
\email[]{grechnev@ilt.kharkov.ua}
\affiliation{B. Verkin Institute for Low Temperature Physics and Engineering,
National Academy of Sciences of Ukraine, 61103 Kharkov, Ukraine}

\author{I.P. Zhuravleva}
\affiliation{B. Verkin Institute for Low Temperature Physics and Engineering,
National Academy of Sciences of Ukraine, 61103 Kharkov, Ukraine}

\author{A.A. Lyogenkaya}
\affiliation{B. Verkin Institute for Low Temperature Physics and Engineering,
National Academy of Sciences of Ukraine, 61103 Kharkov, Ukraine}

\author{V.A. Pashchenko}
\affiliation{B. Verkin Institute for Low Temperature Physics and Engineering,
National Academy of Sciences of Ukraine, 61103 Kharkov, Ukraine}

\author{B.N. Savenko}
\affiliation{Joint Institute for Nuclear Research, 141980 Dubna, Russia}

\author{D. Novoselov}
\affiliation{M.N. Mikheev Institute of Metal Physics,
Ural Branch of the Russian Academy of Sciences, 620108 Yekaterinburg, Russia}
\affiliation{Ural Federal University, 620002 Yekaterinburg, Russia}

\author{D. Prabhakaran}
\affiliation{Department of Physics, University of Oxford, OX1 3PU, Oxford, Great Britain}

\author{I.O. Troyanchuk}
\affiliation{Scientific-Practical Material Research Center,
National Academy of Sciences of Belarus, 220072 Minsk, Belarus}

\begin{abstract}
The effect of pressure on magnetic properties of LaCoO$_3$
is studied experimentally and theoretically.
The pressure dependence of magnetic susceptibility $\chi$ of  LaCoO$_3$ is obtained
by precise measurements of $\chi$ as a function of the hydrostatic pressure
$P$ up to 2 kbar in the temperature range from 78 K to 300 K.
A pronounced magnitude of the pressure effect is found to be negative in sign
and strongly temperature dependent.
The obtained experimental data are analysed by using a two-level model
and DFT+U calculations of the electronic structure of LaCoO$_3$.
In particular, the fixed spin moment method was employed
to obtain a volume dependence of the total energy difference $\Delta$ between the
low spin and the intermediate spin states of LaCoO$_3$.
Analysis of the obtained experimental $\chi(P)$ dependence within the two-level model,
as well as our DFT+U calculations, have revealed the anomalous large decrease
in the energy difference $\Delta$ with increasing of the unit cell volume.
This effect, taking into account a thermal expansion, can be responsible
for the temperatures dependence of $\Delta$,
predicting its vanishing near room temperature.
\end{abstract}

\maketitle
\section{Introduction}
The cobalt oxides are of growing interest since the discovery in them of
a giant magnetoresistance, a large thermoelectric effect, a large value
of the Hall effect, and anomalous magnetic properties.
Their peculiar physical properties partially originate from a rich variety
of the valence and spin states of Co ions.
The Co$^{3+}$ ions in the LaCoO$_3$ compound and in the rare-earth RCoO$_3$ cobaltites
have an electronic 3$d^6$ configuration, and these ions can be in low-spin (LS, $S$=0),
intermediate spin (IS, $S$=1) and high-spin (HS, $S$=2) states.
The energy gap between these states can be quite small and the spin state can vary
with temperature.
This leads to a significant change in electric, magnetic, and transport properties
of RCoO$_3$ compounds \cite{Raveau12,Takami14,Raveau15}.

Lanthanum cobalt oxide LaCoO$_3$ exhibits intriguing magnetic and thermoelectric properties,
also semiconducting and metallic electrical conductivity.
The experimental and theoretical studies of electronic and magnetic properties of
LaCoO$_3$ have been carried out by many groups
(see e.g. Refs.\cite{Raveau12,Takami14,Raveau15} and references therein).
However, a detailed understanding of the magnetic properties observed in
the LaCoO$_3$ compound is still missing.
In the recent years numerous experimental
\cite{Baier05,Kozlenko07,Podlesnyak06,Altarawneh12,Rotter14,Efimov16,Ikeda16,Sikolenko16,Efimov17} 
and theoretical
\cite{Korotin96,Nekrasov03,Knizek05,Spaldin09,Ovchinnikov11,Krapek12,Babkin14,Sotnikov16,Singh17}
studies performed for LaCoO$_3$ have given contradictory results,
and scenario of transition between the magnetic states of Co$^{3+}$ ion with temperature
(low-spin $\Rightarrow$ high-spin or low-spin $\Rightarrow$ intermediate-spin
$\Rightarrow$ high-spin) remains unclear.

To solve this problem, it is necessary to shed more light on the nature of magnetic states
in LaCoO$_3$, which appear to be very sensitive, in particular, to the volume changes.
Therefore, one can expect, that the spin state transitions can be influenced by varying
the pressure.
Actually, the most direct indicator of the spin state of cobalt ions
is the magnetic susceptibility behaviour.
This gives us a direction to investigate the pressure effect on magnetic properties
of LaCoO$_3$ using experimental and theoretical tools in order to elucidate the
mechanism of transitions between the magnetic states of Co$^{3+}$ ions.

The first and apparently the only study of magnetic susceptibility under pressure
in LaCoO$_3$ \cite{Asai97} has revealed a large and strongly temperature-dependent
negative effect.
Later the related results for LaCoO$_3$ were obtained from measurements of
the volume magnetostriction \cite{Sato08}.
However, the comparison of the magnetostriction data \cite{Sato08} with the results
of direct measurements of magnetic susceptibility under pressure \cite{Asai97}
shows their significant quantitative discrepancy, which motivates further study
of magnetovolume effects in this compound.

In this paper, we carried out experimental studies of the influence of hydrostatic
pressure on magnetic susceptibility of LaCoO$_3$ at temperatures from $T=78$ to 300~K.
The obtained experimental data are analyzed by using a two-level model
\cite{Baier05,Zobel02} and DFT+U calculations.
In particular, the fixed spin moment method \cite{Mohn84} was employed
to obtain a volume dependence of the total energy difference between the
LS, IS and HS states of LaCoO$_3$.

\section{Experimental details and results}

\begin{figure}[]
\begin{center}
\includegraphics[width=0.45 \textwidth]{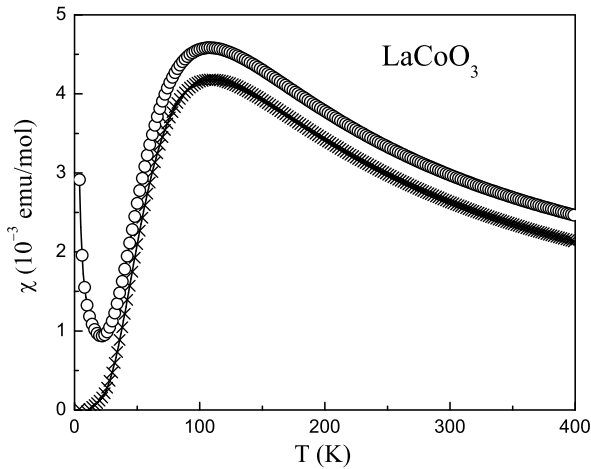}
\caption{Temperature dependence of magnetic susceptibility of LaCoO$_3$:
{\Large$\circ$} -- as measured data, {\large$\times$} -- after subtracting
a low temperature Curie-Weiss term (see text, for details)} \label{X(T)}
\end{center}
\end{figure}

The LaCoO$_3$ single crystal was grown by the floating zone method \cite{Prabhakaran05}
and its single phase and the rhombohedrally distorted perovskite type structure were
confirmed by structural analysis.
For additional characterization of the sample, the temperature dependence of its
magnetic susceptibility $\chi(T)$ was measured in the temperature range of 2--400 K
and a magnetic field of 1 T, using a Quantum Design SQUID magnetometer.
The obtained temperature dependence (Fig. \ref{X(T)}) shows a pronounced maximum
at $T_{\rm max}\simeq$ 105~K and appears to be in agreement with the results
of previous studies \cite{Zobel02,Prabhakaran05,Asai98,Yan04}.
At low temperature, the Curie-Weiss-like contribution was observed,
$\chi_{\rm CW}=C/T+\chi_0$, with $C\simeq10.4\times10^{-3}$ K$\cdot$emu/mole
and $\chi_0\sim0.3\times10^{-3}$ emu/mole.
Here $C/T$ term implies a number of paramagnetic impurities and $\chi_0$ is
the temperature-independent host contribution.
In addition, it should be noted that because of a negligible magnetic anisotropy
in LaCoO$_3$ \cite{Prabhakaran05}, the magnetic field direction can be arbitrarily
chosen relative to the crystallographic axes.

The measurements of the uniform pressure effect on magnetic susceptibility
of LaCoO$_3$ were carried out under helium gas pressure $P$ up to 2 kbar,
using a pendulum type magnetometer \cite{Panfilov15}.
The sample was cut in the shape of parallelepiped with dimensions of
$3\times3\times2$ mm and a mass of 0.125 grams.
It was placed inside a small compensating coil located at
the lower end of the pendulum rod.
Under switching on magnetic field, the value of current through the coil,
at which the magnetometer comes back to its initial position,
is the measure of the sample magnetic moment.
To measure the pressure effects, the pendulum magnetometer was inserted into
a cylindrical non-magnetic pressure cell, which was placed inside a cryostat.
In order to eliminate the effect on susceptibility of the temperature changes
during applying or removing pressure, the measurements were performed at fixed
temperatures in the range between 78 and 300 K.
The relative errors of measurements of $\chi$ under pressure did not exceed 0.1\%
for employed magnetic field $H=1.7$ T.

\begin{figure}[]
\begin{center}
\includegraphics[width=0.4 \textwidth]{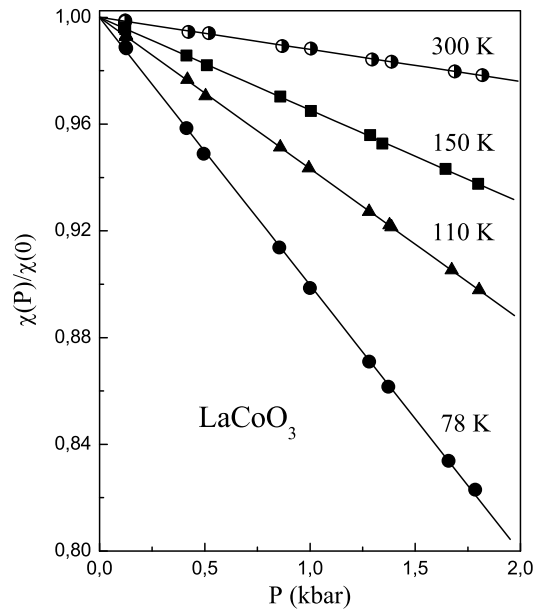}
\caption{Pressure dependence of magnetic susceptibility of LaCoO$_3$,
normalized to its value at $P=0$, at different temperatures} \label{X(P)}
\end{center}
\end{figure}

The experimental $\chi(P)$ dependencies, normalized to the value of $\chi$ at zero pressure,
are presented in Fig. \ref{X(P)}.
As seen in Fig. \ref{X(P)}, there is a pronounced decrease of $\chi$ under pressure,
which is linear within experimental errors.
The corresponding values of the pressure derivative d\,ln$\chi$/d$P$ are listed
in Table  \ref{exp} together with the values of $\chi$ at $P=0$.

\begin{table}[h]
\caption{Magnetic susceptibility $\chi$ of LaCoO$_3$ at $P=0$ and its pressure derivative d\,ln$\chi$/d$P$
at different temperatures.}
\vspace{5pt} \label{exp}
\begin{center}
\begin{tabular}{ccc}
\hline
 T (K)&~~ $\chi$ ($10^{-3}$ emu/mol)~~ &~~ d\,ln$\chi$/d$P$ (Mbar$^{-1}$)\\
\hline
  78 & 4.06  & $ -100 \pm 5$ \\
 110 & 4.36  & $ -56.7 \pm 3$ \\
 150 & 4.12  & $ -34.7 \pm 1.5$ \\
 300 & 2.87  & $ -12\pm 0,5$ \\
\hline
\end{tabular}
\end{center}
\end{table}

\section{Computational details and results}

It has been established, that DFT-LSDA approximation predicts an incorrect
ground state of LaCoO$_3$ \cite{Korzhavyi99,Ravindran02,Harmon13}.
To go beyond the DFT-LSDA, the DFT+U method has been employed and basically provided
the semiconducting ground state of LaCoO$_3$ at ambient conditions
(see e.g. \cite{Korotin96,Nekrasov03,Knizek05,Spaldin09,Singh17}).

The present calculations of volume-dependent electronic structure for LaCoO$_3$
were performed using the linearized augmented plane wave method with
a full potential (FP-LAPW, Elk implementation \cite{elk}).
We also compared the FP-LAPW results on many occasions with the calculations
performed by using the Quantum-Espresso code \cite{Giannozzi09,QE}.
We have used the projector-augmented wave (PAW) potentials \cite{Corso14,Topsakal14},
which are for direct use with the Quantum Espresso code.

The DFT+$U$ approach was employed within the generalized gradient
approximation (GGA) for the exchange-correlation functional \cite{pbe96}.
The effective Coulomb repulsion energy $U_{\rm eff}=U-J=$2.75 eV was adopted
for Co$^{3+}$ ions according to Refs. \cite{Spaldin09,Singh17},
where such value of $U\equiv U_{\rm eff}$ has provided the correct ground-state.

\begin{figure}[]
\begin{center}
\includegraphics[trim=0mm 0mm 0mm 0mm,scale=0.8]{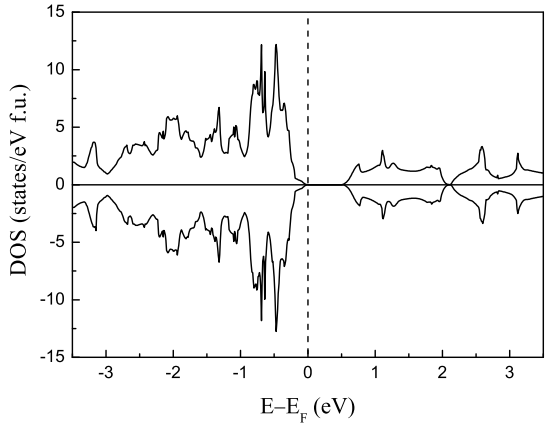}
\caption{Spin resolved density of electronic states for LS configuration
of LaCoO$_3$ [in states/(spin$\cdot$eV)].
The Fermi level is indicated by a vertical dashed line.}
\label{DOS0}
\end{center}
\end{figure}

The calculated density of electronic states (DOS) for the ground state
of LaCoO$_3$ is shown in Fig.~\ref{DOS0}.
Our DFT+U calculations have provided the paramagnetic ground state
with a band gap about 0.5 eV, which is close to the experimental value
\cite{Sharma92}.
For this low-spin state of Co$^{3+}$, the valence band is composed by
the Co $t_{2g}$ states and $2p$ orbitals of oxygen,
whereas the conduction band is formed by the $e_g$ states of Co.
For the low-spin configuration the basic features of calculated here
electronic structure of LaCoO$_3$ are in agreement with results
of previous calculations \cite{Knizek05,Spaldin09,Singh17}.
We have also calculated the volume dependence of the total energy $E(V)$,
and obtained the theoretical equilibrium volume $V_{\rm th}\cong 56$ \AA$^3$
per formula unit of LaCoO$_3$.
This theoretical volume appeared to be slightly larger (about 1.5\%) than
the experimental volume at $T$=5 K (see Ref. \cite{Radaelli02}),
presumably due to the employed GGA+U approach.

\begin{figure}[]
\begin{center}
\includegraphics[trim=0mm 0mm 0mm 0mm,scale=0.8]{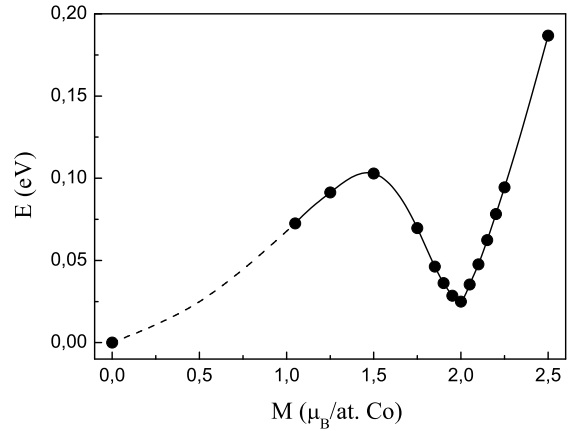}
\caption{Fixed moment calculation of the total energy for LaCoO$_3$
at the theoretical equilibrium volume.
The energies are given relative to the LS state.}
\label{FSM}
\end{center}
\end{figure}

In order to shed light on magnetic properties of LaCoO$_3$
we have employed the fixed spin moment (FSM) method \cite{Mohn84}.
It was demonstrated (see e.g. Ref. \cite{Mohn06}), that FSM method
can provide valuable information about metastable magnetic phases.
By this way we calculated the total energy of LaCoO$_3$ as a function
of the magnetic moment per formula unit.
The results of these fixed spin moment calculations are shown in Fig. \ref{FSM}.

\begin{figure}[]
\begin{center}
\includegraphics[trim=0mm 0mm 0mm 0mm,scale=0.8]{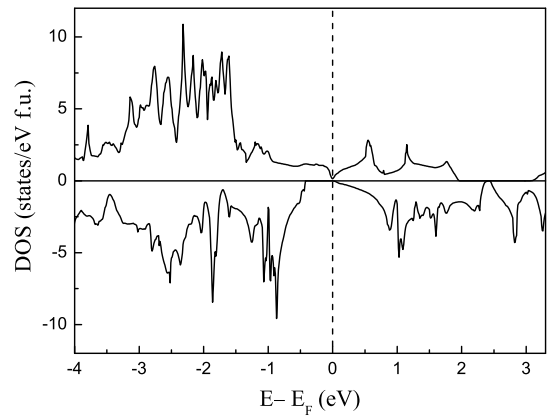}
\caption{Spin resolved density of electronic states for IS configuration
of LaCoO$_3$ [in states/(spin$\cdot$eV)].
The Fermi level is indicated by a vertical dashed line.}
\label{DOS1}
\end{center}
\end{figure}

One can see clear minimum in the $E$ vs. $M$ curve
with magnetic moment of 2 $\mu_{\rm B}$, presumably corresponding to the
intermediate-spin state of Co$^{3+}$ ion in LaCoO$_3$.
According to our FSM calculations, the high-spin state of LaCoO$_3$
($S=2$ of Co$^{3+}$ ion) appeared to be about 1 eV higher in energy
than the LS state.
These results prove that the transition from nonmagnetic to magnetic state
in LaCoO$_3$ actually takes place between LS and IS states.
It is remarkable that the minimum in Fig. \ref{FSM}
is situated energetically only 25 meV above the LS state.
For this IS state of LaCoO$_3$ the spin subbands are split
and partly overlapped, as can be seen in  Fig. \ref{DOS1}.
Therefore our calculated ferromagnetic
IS state appeared to be half metallic, though
the value of DOS at $E_{\rm F}$ is rather small,
and the conduction and valence bands in fact touch each other.

\begin{figure}[]
\begin{center}
\includegraphics[trim=0mm 0mm 0mm 0mm,scale=0.8]{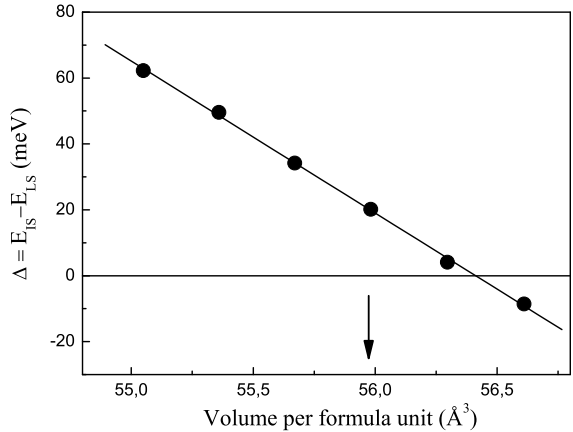}
\caption{Calculated volume dependence of the total energy
difference between the IS and the LS states of LaCoO$_3$.
The arrow indicates the theoretical volume.}
\label{dEV}
\end{center}
\end{figure}

We have also calculated the volume dependence of the total energy
difference between IS and LS states of LaCoO$_3$,
$\Delta = E_{\rm IS}-E_{\rm LS}$,
which is presented in Fig.~\ref{dEV}.
This volume dependence corresponds to substantial increase of $\Delta$ under pressure,
and also indicates a possibility of the spin states crossover under thermal expansion.

\section{Discussion}

The unusual temperature dependence of magnetic susceptibility $\chi(T)$ in LaCoO$_3$
is commonly assumed to be caused by a temperature driven transition of the Co$^{3+}$
ions from the non-magnetic low spin state LS to a magnetic state:
either intermediate spin, IS, or high spin, HS, state.
As was shown in Ref. \cite{Zobel02}, the Co ions contribution to susceptibility,
$\chi_{\rm Co}(T)$, at least up to 300 K, can be successfully described within
LS$\to$IS scenario by the expression for the two-level system \cite{Baier05,Zobel02}
with the energy difference $\Delta$ for these two states:
\begin{equation}
\chi_{\rm Co}(T) = {{N_{\rm A}g^2\mu^2_{\rm B}S(S+1)}\over{3k_{\rm B}T}}\times w(T)\equiv
{C\over T}\times w(T).
\label{X(T)_model}
\end{equation}
Here the multiplier $C/T$ describes the Curie susceptibility of the excited state
supposing a negligible interaction between the Co$^{3+}$ moments;
$N_{\rm A}$ is the Avogadro  number, $\mu_{\rm B}$ the Bohr magneton,
$k_{\rm B}$ the Boltzmann constant, $g$ the Lande factor and $S$ the spin number.
The multiplier $w(T)$ is the population of the excited state expressed by
\begin{equation}
w(T)={{\nu(2S+1){\rm e}^{-\Delta/T}}\over{1+\nu(2S+1){\rm e}^{-\Delta/T}}},
\label{w(T)}
\end{equation}
where $2S+1$ and $\nu$ are the spin and orbital degeneracy of the excited state,
respectively, and the energy difference between the excited and ground states,
$\Delta$, is in units of temperature $T$.

Note that according to the approach used, the behaviour of $\chi_{\rm Co}(T)$ is
determined by the single parameter $\Delta$ and its variation with temperature.
Then dependence $\Delta(T)$, which satisfies the experiment, can be estimated
from Eq. (\ref{w(T)}) as
\begin{equation}
\Delta(T) = T\ln \left[\nu(2S+1){{1-w(T)}\over w(T)}\right].
\label{Delta(T)}
\end{equation}
To analyze the experimental data on $\chi_{\rm Co}(T)$, we used the appropriate values
of the model parameters: $g=2$, $S=1$ and $\nu=1$ (the latter supposes that orbital
degeneracy of IS state is lifted due to local distortions of the crystal lattice).

\begin{figure}[]
\begin{center}
\includegraphics[width=0.45 \textwidth]{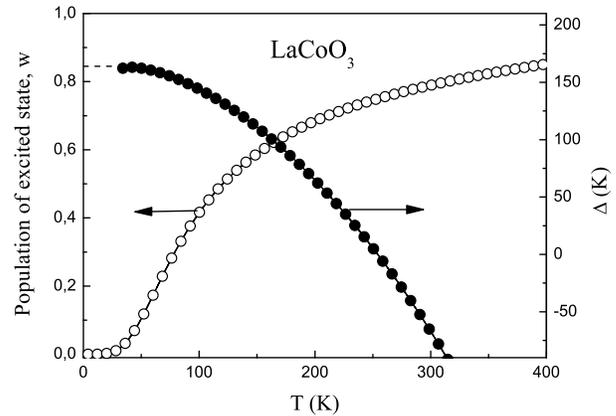}
\caption{Temperature dependencies of the population of excited state, $w(T)$, and
of the energy difference, $\Delta$, between IS and LS states} \label{w'D(T)}
\end{center}
\end{figure}
Using the experimental data on $\chi_{\rm Co}(T)$, we have evaluated within
Eqs. (\ref{X(T)_model}) and (\ref{Delta(T)}) temperature dependencies of
the excited state population, $w(T)$, and of the energy difference $\Delta(T)$,
which are presented in Fig. \ref{w'D(T)}.

It should be noted, that the obtained temperature dependence of the population of excited
state, $w(T)$, appeared to be in a qualitative agreement with the results of
combined analysis of X-ray powder diffraction (XPD) and high-resolution extended
X-ray absorption fine structure (EXAFS) for LaCoO$_3$ in Refs. \cite{Sikolenko16,Efimov17},
and also X-ray magnetic circular dichroism (XMCD) measurements on LaCoO$_3$ single crystal
\cite{Efimov16}.
Specifically, in the temperature range from $\sim 50$ to 300 K the thermally induced
spin-state transition was reported with a gradual growth of the IS state fraction of Co$^{3+}$
ions, while a substantial fraction of Co$^{3+}$ ions remains in the LS state up to temperatures
about 150 K.

As one can see, there is a noticeable decrease in $\Delta(T)$ with increasing temperature
from $\Delta\simeq 165$ K at $T=0$ K to $\Delta=0$ at $T\simeq 250$ K.
The obtained data on temperature dependence of $\Delta$ are close to
that reported in Ref. \cite{Knizek05}.

To analyse the pressure effect data within Eqs. (\ref{X(T)_model}) and (\ref{w(T)}),
we can estimate a pressure derivative of the magnetic susceptibility, d$\chi$/d$P$,
which is given by
\begin{equation}
{{\rm d}\chi(T)\over{\rm d}P}\simeq{{\rm d}\chi_{\rm Co}(T)\over{\rm d}P} = -
\chi_{\rm Co}(T)\left[{1\over T}-{\chi_{\rm Co}(T)\over C}\right]{{\rm d}\Delta\over{\rm d}P} .
\label{dX/dP}
\end{equation}

\begin{figure}[]
\begin{center}
\includegraphics[width=0.46 \textwidth]{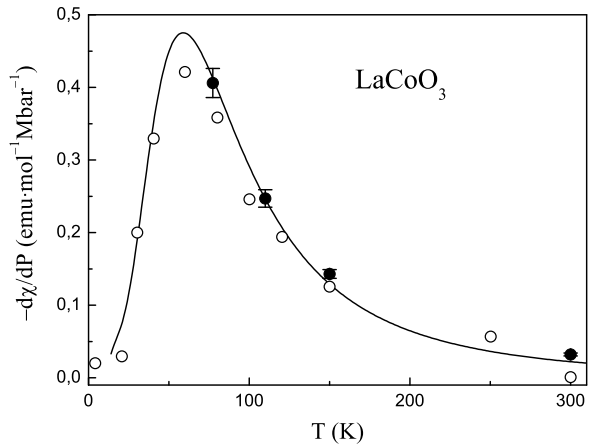}
\caption{Temperature dependence of the pressure derivative d$\chi$/d$P$ in LaCoO$_3$:
({\large$\bullet$}) is the experimental data of the present work,
({\large$\circ$}) the indirect data resulted from volume magnetostriction study \cite{Sato08}.
The solid line is the model description using Eq. (\ref{dX/dP}) (see text, for details)}
\label{dchi-dP}
\end{center}
\end{figure}

The obtained experimental data on d$\chi$/d$P$ are shown in Fig.~\ref{dchi-dP}
and appeared to be in agreement with the related data from measurements
of volume magnetostriction in LaCoO$_3$ \cite{Sato08}.
This set of data can be fairly described by Eq. (\ref{dX/dP})
(solid line in Fig. \ref{dchi-dP}) using the fitting value d$\Delta$/d$P=+12$ K/kbar.

Note that the strong increase of $\Delta$ under pressure implies a substantial
temperature dependence of this parameter due the change in volume via thermal expansion.
The corresponding effect in $\Delta$ can be approximately estimated as:
\begin{equation}
\delta\Delta=\Delta(T)-\Delta(0)\approx{\partial\Delta\over\partial\,{\rm ln}V}\times {{V(T)-V(0)}
\over V(0)},
\label{deltaD(T)}
\end{equation}
where $\partial\Delta/\partial\,{\rm ln}V=-B\times\partial\Delta/\partial P$,
$B$ is the bulk modulus.
Using the values $B\sim 1.35$ Mbar (average value of $B\simeq 1.22$ Mbar \cite{Zhou05}
and 1.5 Mbar \cite{Vogt03}), $(V(300~{\rm K})-V(0))/V(0)\sim 0.015$ \cite{Radaelli02}
and $\partial\Delta/\partial P\simeq 12\times10^3$ K/Mbar (this work),
we obtain a rough estimate of $\delta\Delta$ at $T=300$ K to be about $-240$ K, which
is in a reasonable agreement with $\Delta(T)$ dependence in Fig. \ref{w'D(T)}.

It should be noted that some improvement of the model description can be obtained
by taking into account i) a temperature dependence of elastic properties \cite{Naing06},
ii) a possible manifestation of the interaction between Co$^{3+}$ moments
in the excited IS state and iii) a contribution of HS state at higher temperatures range,
which are not considered in the present analysis.
Nevertheless, we presume that improvements of the model will not lead
to noticeable changes in the obtained parameters:
\begin{equation}
\Delta\simeq165~{\rm K~at}~T= 0~{\rm K, ~~~ d}\Delta/{\rm d}P\simeq 12~{\rm K/kbar}.
\end{equation}

\begin{figure}[]
\begin{center}
\includegraphics[width=0.45 \textwidth]{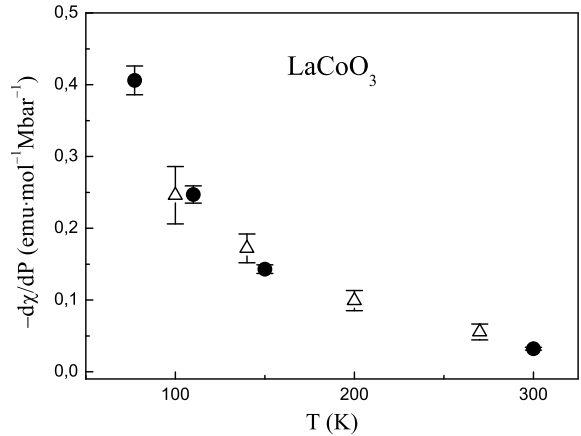}
\caption{Temperature dependence of the pressure derivative d$\chi$/d$P$ in LaCoO$_3$:
({\large$\bullet$}) are the experimental data on the hydrostatic pressure effect of this
work, ({$\triangle$}) the estimated by Eq. (\ref{dX/dP_chemical}) the chemical pressure effect
(see text, for details)} \label{dchi-dP_common}
\end{center}
\end{figure}

We note that our estimate of d$\Delta$/d$P$ appeared to be close to
the reported value of the chemical pressure effect on $\Delta$,
d$\Delta$/d$P\sim 10$ K/kbar, resulted from analysis of the magnetic properties
of La$_{1-x}$Pr$_x$CoO$_3$ system at low concentrations $x$ \cite{Kobayashi06}.
In addition, the experimental results on concentration dependencies
of Co$^{3+}$ ions contribution to $\chi$, $\chi_{\rm Co}(x)$, and lattice cell volume, $V(x)$,
for the iso-structural compounds La$_{1-x}$Pr$_x$CoO$_3$ ($0\leq x\leq 0.3$) \cite{Kobayashi06}
allow to evaluate the chemical pressure effect on magnetic susceptibility of LaCoO$_3$ as:
\begin{equation}
{{\rm d}\chi(T)\over{\rm d}P}\simeq{{\rm d}\chi_{\rm Co}(T)\over{\rm d}P}=
{\partial\chi_{\rm Co}(T)\over\partial x}\left(-B{\partial\,{\rm ln}V\over
\partial x}\right)^{-1}.
\label{dX/dP_chemical}
\end{equation}
Here we used the experimental data on $\chi_{\rm Co}(T)$, the value
$\partial\,{\rm ln}V/\partial x\simeq -0.03$ of Ref. \cite{Kobayashi06}
and the bulk modulus value $B=1.35\pm 0.15$ Mbar.
The resulting estimates of the derivative of magnetic susceptibility with respect
to the chemical pressure for arbitrarily chosen temperatures in the range $78-300$ K
are shown in Fig. \ref{dchi-dP_common}.
As is seen, these estimates reasonably agree with the experimental data
on the hydrostatic pressure effects in LaCoO$_3$.

Finally it should be noted, that theoretical studies of the volume dependence of
$\Delta$ also show a positive pressure effect.
However the estimated from data of Ref.~\cite{Korotin96} values of both d$\Delta$/d$P$
and $\Delta$ itself appear to be an order of magnitude higher than the experimental
results of the present work.
On the other hand, our DFT+U calculations have provided the values
$\Delta(0)\simeq 230$ K and  d$\Delta$/d$P\simeq 22$ K/kbar, which appeared to be in
much better agreement with that obtained from the present experiments.

As it has been demonstrated, the experimental data on the pressure effect in magnetic
susceptibility of LaCoO$_3$, as well as its $\chi(T)$ dependence at ambient pressure,
are satisfactorily described at low and moderate temperatures within
LS$\to$IS scenario by the two-level system model with energy splitting $\Delta$.
This gives an evidence of the applicability of the approach used to describe
magnetic properties of LaCoO$_3$ and related RCoO$_3$ compounds
at least up to room temperature.

\section{Concluding remarks}

In summary, the results of our study of the hydrostatic pressure effects on magnetic
susceptibility of LaCoO$_3$, combined with the related literature data on volume
magnetostriction  \cite{Sato08}, have been consistently described within
the approach of a thermal population of the IS state of the Co$^{3+}$ ions.
We analysed our data using a simple two-level model \cite{Baier05,Zobel02}
and revealed the anomalous large increase in the energy difference
$\Delta$ between LS and IS states with increasing pressure.

The estimated magnitude of this effect is in a reasonable agreement with the
results of present DFT+U calculations of the volume dependence of $\Delta$.
Our results are also consistent with the literature data on manifestation of
the pressure-induced continuous depopulation of the IS state in the behaviour
of the physical properties of LaCoO$_3$ under high pressures
(see Refs. \cite{Vogt03,Vanko06}).

In addition, the established similarity of the effects of physical and chemical 
pressures on $\Delta$ and $\chi$ allows to conclude that the spin state of Co$^{3+}$
ions in LaCoO$_3$ and related RCoO$_3$ cobaltites is predominantly governed by
the interatomic spacing variations.

Finally, it should be noted that in recent works
(see e.g. \cite{Podlesnyak06,Altarawneh12,Rotter14,Ikeda16,Krapek12,Babkin14})
there were attempts to revive the LS--HS scenario,
and the problem of the true spin state transitions in LaCoO$_3$ is still
a subject of debate.
In this connection, however, one should take into consideration the results of
recent XPD and EXAFS diffraction studies \cite{Sikolenko16,Efimov17} on
polycrystalline LaCoO$_3$ samples.
It was found that certain amount of Co$^{3+}$ ions in the HS state are located
predominantly within the surface layer of the crystallines, and this effect can
be explained by influence of structural defects due to oxygen vacancies and
distorted chemical bonds at the boundary of powder grains.

\section{Acknowledgement}
This work was supported by the Russian Foundation for Basic Research
according to the research project 17-32-50042-mol\_nr.


\begin{thebibliography}{99}

\bibitem{Raveau12} B. Raveau and M. Seikh, {\em Cobalt Oxides: From Crystal Chemistry to Physics},
Wiley-VCH, Weinheim (2012).
http://dx.doi.org/10.1002/9783527645527


\bibitem{Takami14} T. Takami, {\em Functional Cobalt Oxides: fundamentals, properties, and
applications}, Pan Stanford Publishing, Singapore (2014).

\bibitem{Raveau15} B. Raveau and M. Seikh, in: K.H.J. Buschow (Ed), {\em Handbook of Magnetic
Materials}, North Holland, Amsterdam, Vol. {\bf 23}, chap. 3, 161 (2015).\\
http://dx.doi.org/10.1016/B978-0-444-63528-0.00003-6


\bibitem{Baier05} J. Baier, S. Jodlauk, M. Kriener, A. Reichl, C. Zobel, H. Kierspel,
A. Freimuth, and T. Lorenz, {\em Phys. Rev. B} {\bf 71}, 014443 (2005).\\
http://dx.doi.org/10.1103/PhysRevB.71.014443


\bibitem{Podlesnyak06}A. Podlesnyak, S. Streule, J. Mesot, M. Medarde, E. Pomjakushina,
K. Conder, A. Tanaka, M.W. Haverkort, and D.I. Khomskii,
{\em Phys. Rev. Lett.}  {\bf 97}, 247208 (2006).\\
http://dx.doi.org/10.1103/PhysRevLett.97.247208


\bibitem{Kozlenko07} D.P. Kozlenko, N.O. Golosova, Z. Jir\'ak, L.S. Dubrovinsky,
B.N. Savenko, M.G. Tucker, Y. Le Godec, and V.P. Glazkov,
{\em Phys. Rev. B} {\bf 75}, 064422 (2007). \\
http://dx.doi.org/10.1103/PhysRevB.75.064422


\bibitem{Altarawneh12} M.M. Altarawneh, G.-W. Chern, N. Harrison, C.D. Batista, A. Uchida,
M. Jaime, D.G. Rickel, S.A. Crooker, C.H. Mielke, J.B. Betts, J.F. Mitchell, and M.J.R. Hoch,
{\em Phys. Rev. Lett.} {\bf 109}, 037201 (2012).\\
http://dx.doi.org/10.1103/PhysRevLett.109.037201


\bibitem{Rotter14} M. Rotter, Z.-S. Wang, A.T. Boothroyd, D. Prabhakaran, A. Tanaka, and
M. Doerr, {\em Scientific Reports} {\bf 4}, 7003 (2014).\\
http://dx.doi.org/10.1038/srep07003


\bibitem{Efimov16}V. Efimov, A. Ignatov, I.O. Troyanchuk, V.V. Sikolenko,
A. Rogalev, F. Wilhelm, E. Efimova, S.I. Tiutiunnikov, D. Karpinsky,
V. Kriventsov, E. Yakimchuk, S. Molodtsov, P. Sainctavit and D. Prabhakaran,
{\em J. Phys. Conf. Series} {\bf 712}, 012111 (2016).\\
http://dx.doi.org/:10.1088/1742-6596/712/1/012111


\bibitem{Ikeda16} A. Ikeda, T. Nomura, Y.H. Matsuda, A. Matsuo, K. Kindo, and K. Sato,
{\em Phys. Rev. B} {\bf 93}, 220401(R) (2016).\\
http://dx.doi.org/10.1103/PhysRevB.93.220401


\bibitem{Sikolenko16}V.V. Sikolenko, I.O. Troyanchuk, V.V. Efimov,
E.A. Efimova, S.I. Tiutiunnikov, D.V. Karpinsky, S. Pascarelli, O. Zaharko,
A. Ignatov, D. Aquilanti, A.G. Selutin, A.N. Shmakov and D. Prabhakaran,
{\em J. Phys. Conf. Series} {\bf 712}, 012118 (2016).\\
http://dx.doi.org/10.1088/1742-6596/712/1/012118


\bibitem{Efimov17} V. Efimov, V. Sikolenko, I.O. Troyanchuk, D. Karpinsky, E. Efimova,
S.I. Tiutiunnikov, B.N. Savenko, D. Novoselov, and D. Prabhakaran,
{\em Powder Diffraction} {\bf 32}, S151 (2017).\\
http://dx.doi.org/:10.1017/S0885715617000148

\bibitem{Korotin96} M.A. Korotin, S.Yu. Ezhov, I.V. Solovyev, V.I. Anisimov, D.I. Khomskii,
and G.A. Sawatzky, {\em Phys. Rev. B} {\bf 54}, 5309 (1996).\\
http://dx.doi.org/10.1103/PhysRevB.54.5309

\bibitem{Nekrasov03} I.A. Nekrasov, S.V. Streltsov, M.A. Korotin, and V.I. Anisimov,
{\em Phys. Rev. B} {\bf 68}, 235113 (2003).\\
http://dx.doi.org/ 10.1103/PhysRevB.68.235113


\bibitem{Knizek05} K. Kn\'{i}\v{z}ek, P. Nov\'ak, and Z. Jir\'ak,
{\em Phys. Rev. B} {\bf 71}, 054420 (2005).\\
http://dx.doi.org/10.1103/PhysRevB.71.054420


\bibitem{Spaldin09} J.M. Rondinelli and N.A. Spaldin, {\em Phys. Rev. B} {\bf 79}, 054409 (2009).\\
http://dx.doi.org/10.1103/PhysRevB.79.054409


\bibitem{Ovchinnikov11} S.G. Ovchinnikov, Yu.S. Orlov, I.A. Nekrasov, and V. Pchelkina,
{\em J. Exp. Theor. Phys.} {\bf 112}, 140 (2011).\\
http://dx.doi.org/10.1134/S1063776110061159


\bibitem{Krapek12} V. K\v{r}\'{a}pek, P. Nov\'{a}k, J. Kune\v{s}, D. Novoselov,
Dm. M. Korotin, and V. I. Anisimov, {\em Phys. Rev. B} {\bf 86}, 195104 (2012).\\
http://dx.doi.org/10.1103/PhysRevB.86.195104


\bibitem{Babkin14} R.Yu. Babkin, K.V. Lamonova, S.M. Orel, S.G. Ovchinnikov,
and Yu.G. Pashkevich, {\em J. Exp. Theor. Phys. Lett.} {\bf 99}, 476 (2014).\\
http://dx.doi.org/10.1134/S0021364014080037


\bibitem{Sotnikov16}A. Sotnikov and  J. Kune\v{s}, {\em Scientific Reports} {\bf 6}, 30510 (2016).\\
http://dx.doi.org/10.1038/srep30510


\bibitem{Singh17} S. Singh and S.K. Pandey, {\em Philosophical Magazine B} {\bf 97}, 451 (2016).\\
http://dx.doi.org/10.1080/14786435.2016.1263404


\bibitem{Asai97} K. Asai, O. Yokokura, M. Suzuki, T. Naka, T. Matsumoto, H. Takahashi,
N. M\^ori and K Kohn, {\em J. Phys. Soc. Japan}  {\bf 66}, 967 (1997).\\
http://dx.doi.org/10.1143/JPSJ.66.967


\bibitem{Sato08} K. Sato, M.I. Bartashevich, T. Goto, Y. Kobayashi, M. Suzuki, K. Asai, A. Matsuo, and K. Kindo, {\em J. Phys. Soc. Japan} {\bf 77}, 024601 (2008).\\
http://dx.doi.org/10.1143/JPSJ.77.024601


\bibitem{Zobel02} C. Zobel, M. Kriener, D. Bruns, J. Baier, M. Gr\"uninger, T. Lorenz,
P. Reutler and A. Revcolevschi, {\em Phys. Rev. B} {\bf 66}, 020402(R) (2002).\\
http://dx.doi.org/10.1103/PhysRevB.66.020402


\bibitem{Mohn84}K. Schwarz and P. Mohn, {\em  J. Phys. F: Metal Phys.} {\bf 14}, L129 (1984);
V.L. Moruzzi, P.M. Marcus, K. Schwarz, and P. Mohn, {\em Phys. Rev. B} {\bf 34}, 1784 (1986).\\
http://dx.doi.org/10.1103/PhysRevB.34.17


\bibitem{Prabhakaran05} D. Prabhakaran, A.T. Boothroyd, F.R. Wondre, and T.J. Prior,
{\em J. Crystal Growth} {\bf 275}, e827 (2005).\\
http://dx.doi.org/

\bibitem{Asai98} K. Asai, A. Yoneda, O Yokokura, J.M. Tranquada, G. Shirane and K Kohn, {\em J. Phys. Soc. Japan} {\bf 67}, 290 (1998).\\
http://dx.doi.org/10.1143/JPSJ.67.290


\bibitem{Yan04} J.-Q. Yan, J.-S. Zhou, and J.B. Goodenough,
{\em Phys. Rev. B} {\bf 69}, 134409 (2004).\\
http://dx.doi.org/10.1103/PhysRevB.69.134409


\bibitem{Panfilov15} A.S. Panfilov, {\em Low. Temp. Phys.} {\bf 41}, 1029 (2015).\\
http://dx.doi.org/10.1063/1.4938094


\bibitem{Korzhavyi99}P. Ravindran, P.A. Korzhavyi, H. Fjellvag, and A. Kjekshus,
{\em Phys. Rev. B} {\bf 60}, 16423 (1999).\\
http://dx.doi.org/10.1103/PhysRevB.60.16423


\bibitem{Ravindran02}P. Ravindran,  H. Fjellvag, A. Kjekshus, P. Blaha, K. Schwarz, and J. Luitz,
 {\em J. Appl. Phys.} {\bf 91}, 291 (2002).\\
http://dx.doi.org/10.1063/1.1418001


\bibitem{Harmon13} Y. Lee and B.N. Harmon, {\em J. Appl. Phys.} {\bf 113}, 17E145 (2013).\\
http://dx.doi.org/10.1063/1.4798350


\bibitem{elk} http://elk.sourceforge.net/

\bibitem{Giannozzi09}P. Giannozzi, et. al., {\em J.Phys.:Condens.Matter} {\bf 21}, 395502 (2009).\\
http://dx.doi.org/10.1088/0953-8984/21/39/39502


\bibitem{QE} http://www.quantum-espresso.org/

\bibitem{Corso14}A. Dal Corso, {\em Comput. Mater. Sci.} {\bf 95}, 337 (2014).
http://theossrv1.epfl.ch/Main/Pseudopotentials\\
http://dx.doi.org/10.1016/j.commatsci.2014.07.043

\bibitem{Topsakal14} M. Topsakal and R.M. Wentzcovitch, {\em Comput. Mater. Sci. \bf 95} 263 (2014).

http://vlab.msi.umn.edu/resources/repaw\\
http://dx.doi.org/10.1016/j.commatsci.2014.07.030


\bibitem{pbe96} J.P. Perdew, K. Burke, and M. Ernzerhof, {\em Phys. Rev. Lett.}
{\bf 77}, 3865 (1996).\\
http://dx.doi.org/10.1103/PhysRevLett.77.3865


\bibitem{Sharma92}A. Chainani, M. Mathew, and D.D. Sharma,
{\em Phys. Rev. B} {\bf 46}, 9976 (1992).\\
http://dx.doi.org/10.1103/PhysRevB.46.9976


\bibitem{Radaelli02} P.G. Radaelli and S.-W. Cheong, {\em Phys. Rev. B} {\bf 66}, 094408 (2002).\\
http://dx.doi.org/10.1103/PhysRevB.66.094408


\bibitem{Mohn06} M. Sieberer, S. Khmelevskyi, and P. Mohn, {\em Phys. Rev. B}
{\bf 74}, 014416 (2006).\\
http://dx.doi.org/10.1103/PhysRevB.74.014416


\bibitem{Zhou05} J.-S. Zhou, J.-Q. Yan, and J.B. Goodenough,
{\em Phys. Rev. B} {\bf 71}, 220103(R) (2005).\\
http://dx.doi.org/10.1103/PhysRevB.71.220103


\bibitem{Vogt03} T. Vogt, J.A. Hriljac, N.C. Hyatt, and P. Woodward,
{\em Phys. Rev. B} {\bf 67}, 140401(R) (2003).\\
http://dx.doi.org/10.1103/PhysRevB.67.140401


\bibitem{Kobayashi06} Y. Kobayashi, T. Mogi, and K. Asai,
{\em J. Phys. Soc. Japan} {\bf 75}, 104703 (2006).\\
http://dx.doi.org/10.1143/JPSJ.75.104703


\bibitem{Naing06} T.S. Naing, T. Kobayashi, Y. Kobayashi, M. Suzuki and K. Asai,
{\em J. Phys. Soc. Japan} {\bf 75}, 084601 (2006).\\
http://dx.doi.org/10.1143/JPSJ.75.084601


\bibitem{Vanko06} G. Vank\'o, J.-P. Rueff, A. Mattila, Z. N\'emeth, and A. Shukla,
{\em Phys. Rev. B}  {\bf 73}, 024424 (2006).\\
http://dx.doi.org/10.1103/PhysRevB.73.024424

\end{thebibliography}
\end{document}